\documentstyle[12pt]{article}
\begin{document}

\begin{titlepage}

\begin{center}
{\Large {\bf Electroweak characteristics of mesons in the relativistic quark
model}}
{\large V.V.Andreev, N.V.Maksimenko
Gomel State University, Physics Department,\\
Gomel, 246699 Belarus\\
E-mail: andreev@gsu.unibel.by\\}
\end{center}
$$
$$
\begin{center}
{\large Published in Proceedings IV International School-Seminar \\
"The actual problems of particle physics"\\
(Belarus,Gomel,1997)\\
Dubna. 1998,   E1,2-98-307, Vol.1. p. 270-283\\
}
\end{center}
$$
$$
$$
$$
\begin{center}
{\Huge Abstract\\}
\end{center}
\noindent
\begin{center}
{\Large
In this work the electromagnetic radius and the polarizability of the mesons
are obtained by use of the effective Lagrangians
constructed on the one hand with taking into account of general principles
of the relativistic quantum field theory and on the other hand with taking
into account of the compound relativistic model meson representation. Also
using a relativistic constituent quark model based on point form of
Poincare-covariant quantum mechanics we calculated electromagnetic radius
and the decay constants of the mesons with spinor quarks.
}
\end{center}
\vspace*{3.0mm}
\end{titlepage}

\newpage
\begin{center}
{\Large {\bf Electroweak characteristics of mesons in the relativistic quark
model}}
{\large V.V.Andreev, N.V.Maksimenko
Gomel State University, Physics Department,\\
Gomel, 246699 Belarus\\
E-mail: andreev@gsu.unibel.by\\}
$$
$$
{\bf Introduction}
\end{center}

Relativistic few body problem has received a great attention in hadronic and
nuclear physics. The most complete results here exist for the case of two
particles. Description of the bound system in the relativistic quantum field
theory is founded on the four-dimensional covariant Bethe-Salpeter equation
\cite{BeteS}. However, this equation gives series of difficulties when the
practical calculations are made.

There exist various reductions of the two-body Bethe-Salpeter equation.
Different forms of this reduction were discussed Logunov-Tavkhelidze \cite
{LogTav}, Kadyshevsky \cite{Kad1}, Todorov \cite{Tod}, Gross \cite{Gross},
Poluzou,Keister,Lev \cite{Keist1}-\cite{Lev4} and many others.

The relativistic two-body systems was analyzed by Weinberg \cite{Wein},
Frankfurt - Strikman \cite{FrS}, Kondratyuk-Terent'ev \cite{KonTer} in
infinite-momentum frame. Some authors use diagrammatic approach i.e. they
select leading diagrams and project them onto the three-dimensional space.
Others make use effective Hamiltonians, Faddev equation.

A more general covariant perturbation theory with spurions was proposed
Kadyshevsky \cite{Kad2}. Different forms of the quasi-potential equation can
be derived using this approach \cite{Kad2}-\cite{Kad3}. The method of the
quasipotential approach and the method of the contact interaction approach
as consequences of the Bethe-Salpeter equation obtained broad recognition in
the practical application \cite{m2}.

Last years accuracy of measurement of the hadron electromagnetic
characteristic improves considerably. This fact permits to analyze the
existing field-theoretical and model conception of interaction of the
hadrons with the electromagnetic field do better: which why we investigate
some electroweak properties of the mesons.

In the paper \cite{m3} in the framework of the quasipotential approach \cite
{LogTav}, relativistic composite models for computation of the electric
polarizability and the mean square radius of charge distribution and radial
excitations of mesons as bound systems of valence quarks are proposed. In
this work the electromagnetic radius and the polarizability of the mesons
are obtained by use of the effective Lagrangians \cite{m5}-\cite{m7},
constructed on the one hand with taking into account of general principles
of the relativistic quantum field theory and on the other hand with taking
into account of the compound relativistic model meson representation. Also
using a relativistic constituent quark model based on point form of
Poincare-covariant quantum mechanics we calculated electromagnetic radius
and the decay constants of the mesons with spinor quarks.

\begin{center}
{\bf Effective Lagrangian approach }
\end{center}

Consider the scalar bound system which consists of the charged scalar
particles. The effective Lagrangian of interaction of the electromagnetic
field with the hadrons is represented as covariant expansion in terms of
tensor of the electromagnetic field $F_{\mu \nu }$ and potential $A_\mu $
\[
L_{eff}=\frac 1{8M^2}\left( \partial _\mu \varphi ^{+}\right) \left(
\partial _\nu \varphi \right) +\left( \partial _\mu \varphi \right) \left(
\partial _\nu \varphi ^{+}\right) -
\]
\[
-\left( \varphi ^{+}\partial _\mu \partial _\nu \varphi +\varphi \partial
_\mu \partial _\nu \varphi ^{+}\right) \left[ \left( \alpha +\beta \right)
F^{\mu \sigma }F_\sigma ^\nu -\frac 12\beta g_{\mu \nu }F^{\rho \sigma
}F_{\sigma \rho }\right] +
\]
\begin{equation}
+e^2A_\mu ^2\varphi ^{+}\varphi +ie\left[ \varphi ^{+}\stackrel{
\leftrightarrow }{\partial _\mu }\varphi \right] A^\mu ,  \label{m1}
\end{equation}
where $\varphi $ is the field function of scalar charge structural particle,
$\alpha $ and $\beta $ is the electric and magnetic static polarizabilities.
Let us consider a generating functional
\begin{equation}
exp(iS_{eff})=Z(A).  \label{m2}
\end{equation}
For calculation $S_{eff}$ in (\ref{m2}) we confine ourselves by the model
with two the vertex $\Gamma $. As results we have
\begin{equation}
iS_{eff}=\langle \Delta (A;Z_1-Z_1^{^{\prime }})\Gamma (A;Z_1^{^{\prime
}},Z_2^{^{\prime }};Z)\Delta (A;Z_2-Z_2^{^{\prime }})\Gamma
(A;Z_1,Z_2;Z)\rangle ,  \label{m3}
\end{equation}
where $\Delta (A;Z_1-Z_2)$ is the scalar particle propagator in
electromagnetic field, $\Gamma (A;Z_1,Z_2;Z)$ is the three-point vertex
function in the electromagnetic field. Using the reduction technique
methods, one can get a connection of the Bethe-Salpeter function with the
vertex function
\begin{equation}
\chi _{{\cal P}}\left( Z_{1,}Z_2\right) =i\left\langle \exp \left( -i{\cal P}
Z\right) \Delta \left( Z_1-Z_1^{\prime }\right) \Gamma \left( Z_1^{\prime
},Z_2^{\prime };Z\right) \Delta \left( Z_2^{\prime }-Z_2\right)
\right\rangle ,  \label{m4}
\end{equation}
where ${\cal P}$ is the total impulse of compound system. The brackets in
this equation mean the integration with respect to $Z_1^{^{\prime
}},Z_2^{^{\prime }}$ and $Z$ variables. The function (\ref{m4}) obeys to the
homogeneous Bethe-Salpeter equation ~\cite{m8}~.
\begin{equation}
\chi _{{\cal P}}(Z_1,Z_2)=i\langle \Delta (Z_1-Z_1^{^{\prime }})\chi _{{\cal
P}}(Z_1^{^{\prime }},Z_2^{^{\prime }})\Delta (Z_2^{^{\prime
}}-Z_2)V(Z_1^{^{\prime }}-Z_2^{^{\prime }})\rangle ,  \label{m5}
\end{equation}
where $V$ is the interaction potential. In the contact interaction
approximation and in impulse representation the equation (\ref{m5}) is:
\begin{equation}
\Delta _1^{-1}(p_1)\Delta _2^{-1}(p_2)\chi _{{\cal P}}(p)=i\frac g{(2\pi
)^4}\int dq\chi _{{\cal P}}(q),  \label{m6}
\end{equation}
where $p=p_1+p_2,{\cal P}=\frac 12(p_1-p_2).$ After integration over $dp_0$
in (\ref{m6}) we obtain
\begin{equation}
\chi _{{\cal P}}(\vec{p})=N\frac{E_1+E_2}{E_1E_2[(E_1+E_2)^2-E^2]},
\label{m7}
\end{equation}
where $E_i=\sqrt{{\vec{p}}_i^2+m_i^2};~i=1,2;~E=\sqrt{{\vec{{\cal P}}}^2+M^2}
$, $N$ is the normalizing factor, $M$ is the compound system mass. Using the
relativistic covariant normalization ~\cite{m9,m10}~.
\begin{equation}
\frac i{(2\pi )^4}\int dpdp^{^{\prime }}\chi _{{\cal P}}(p)[\frac \partial
{\partial E}(K-V)]\chi _{{\cal P}}(p^{^{\prime }})=2E,  \label{m8}
\end{equation}
where
\[
K(p,p^{^{\prime }};{\cal P})=\Delta _1^{-1}(p_1)\Delta _2^{-1}(p_2)\delta
(p-p^{^{\prime }}),
\]
we obtain
\[
N^2=\frac 1{(\frac{2m^2}{M\beta }atan\frac M{2\beta }-1)},
\]
where $\beta =\sqrt{m^2-\frac{M^2}4}.$

\begin{center}
{\bf Electromagnetic radius and the polarizabilities .}
\end{center}

The pion $\langle r^2\rangle $ is calculated, using (\ref{m7}), is
\begin{equation}
\langle r^2\rangle =\frac 1{8\beta ^2M^2}\frac{M^2+\frac{M^4+48\beta
^4}{2M\beta } atan \frac M{2\beta }-12\beta ^2}{\frac{2m^2}{
M\beta }atan\frac M{2\beta }-1}.  \label{m9}
\end{equation}
It follows from (\ref{m9}), when the connection energy is weak (the compound
system is non-relativistic)
\begin{equation}
\langle r^2\rangle \sim \frac 1{8\beta ^2},  \label{m10}
\end{equation}
and when the connection energy is strong (the compound system is
relativistic)
\begin{equation}
\langle r^2\rangle \sim \frac{0.3}{m^2}.  \label{m11}
\end{equation}
The numerical analyze (\ref{m9}) gives evidence of accordance with the
experimental value $\langle r^2\rangle $ ~\cite{m11}~ when the mass of a $
\pi $-meson is $M=140$ $MeV$ and the mass of a quark is $m=210$ $MeV$. This
result is in accordance with (\ref{m11}). Consequently, $\pi $-meson is the
strong-bound relativistic quark system. In order to calculate the electrical
polarizability of a $\pi $-meson we confined by the model of interaction of
the electromagnetic field. It follows from the equations (\ref{m1}), (\ref
{m3}) and (\ref{m7}) the electrical polarizability of a $\pi $-meson is
defined by
\begin{equation}
\alpha =(\frac{e^2}{4\pi })\frac{16\pi N^2}{M^3}{\cal I}.  \label{m12}
\end{equation}
Here
\[
{\cal I}=\int\limits_0^1dx\{(\widehat{Q}^2+\widehat{\bar Q}^2)\frac{\frac
14x^2-\frac 13x^3+\frac 18x^4}{[x(1-x)-\eta ^2]^2}-\widehat{Q}\widehat{
\bar Q}\frac{\frac 14x^2(1-x)^2}{[x(1-x)-\eta ^2]^2}-
\]
\begin{equation}
-(\widehat{Q}^2+\widehat{\bar Q}^2)\frac{x^3(1-x)^2(\frac 23-\frac 12x)}{
[x(1-x)-\eta ^2]^3}+\widehat{Q}\widehat{\bar Q}\frac{x^3(1-x)^3}{
[x(1-x)-\eta ^2]^3}\},  \label{m13}
\end{equation}
where $\widehat{Q}$ and $\widehat{\bar Q}$ are the values of quark and
antiquark charges, $\eta =\frac mM.$ In the strong coupling we obtain from $
E_q$ (\ref{m13}).
\begin{equation}
\alpha =\frac{e^2}{4\pi }\frac{96\pi }{m^2M}\{(\widehat{Q}^2+\widehat{\bar Q}
^2)(\frac 1{40}+\frac{2M^2}{315m^2})-(\widehat{Q}\widehat{\bar Q}
)(\frac 1{120}+\frac{M^2}{140m^2})\}.  \label{m14}
\end{equation}
The numerical estimations (\ref{m14}), when $m=210\leq MeV$, have a value $
\alpha $, which agrees to an experiment ~\cite{m12}~ we shell explain how
can be calculated the polarizability in the quasipotential approach ~\cite
{LogTav}~. Let us assume that
\[
\Gamma (A|Z_1Z_2Z)=\widetilde{\Gamma }(A)\delta (Z_{10}-Z_0)\delta
(Z_{20}-Z_0).
\]
Then one can derive the equation
\begin{equation}
iS_{eff}=\langle \widetilde{\Gamma }(A)\widetilde{{\cal G}}^{(0)}(A)
\widetilde{\Gamma }(A)\rangle ,  \label{m15}
\end{equation}
where $\widetilde{{\cal G}}^{(0)}(A)$ is the three-dimensional Green
function in the external electromagnetic field. Let us introduce the wave
function, that will obey the quasipotential equation
\begin{equation}
\widetilde{{\cal G}}^{(0)-1}(A)\widetilde{\Phi }(A)=\widetilde{V}(A)
\widetilde{\Phi }(A),  \label{m16}
\end{equation}
where $\widetilde{V}(A)$ is the quasipotential of a system in the
electromagnetic field. Then (\ref{m15}) takes the form
\begin{equation}
iS_{eff}=\langle \widetilde{\Phi }(A)\widetilde{{\cal G}}^{(0)-1}(A)
\widetilde{\Phi }(A)\rangle .  \label{m17}
\end{equation}
Consider a particular case when the system interacts with the constant
electric field. Restricting by the first order in field $\vec E$ in the
left- and right-hand side (\ref{m17}) we get the electric radius of a system
\begin{equation}
\langle r^2\rangle =\int d\vec q\varphi _n(\vec q)(-\frac i2\vec \partial
_q)^2\varphi _n(\vec q)+\frac 3{16}\int d\vec q{\mid \varphi _n(\vec q)\mid }
^2\frac{{\vec q}^2}{E_q^4},  \label{m18}
\end{equation}
where $\varphi _n(\vec q)$ is the wave functions (\ref{m18}) of the relative
movement of the quark in the rest frame system, $E_q=\sqrt{{\vec q}^2+m^2}$,
$m$ is the quark mass. Comparing the structures on (\ref{m17}) being
proportional to the second order in field $\vec E$ we obtain the expression
for the electric polarizability ~\cite{m3}~
\[
\alpha =\frac 1{{\mid \vec E\mid }^2}\{\sum_\xi \int d\vec q\varphi _n(\vec
q)\chi _n^{+}(\xi )i(\widehat{Q_1}-\widehat{Q_2})(\vec E\vec \partial
_q)\varphi _n^{(1)}(\vec q,\xi )-
\]
\[
-\frac 1M\sum_\xi \int d\vec q\varphi _n(\vec q)\chi _n^{+}(\xi )\Gamma
^{(2)}(\vec q)\chi _n(\xi )\varphi _n(\vec q)-
\]
\begin{equation}
-\frac 1M\sum_\xi \int d\vec q\varphi _n(\vec q)\chi _n^{+}(\xi )[\frac 12(
\widehat{Q_1}-\widehat{Q_2})(\vec E\vec \partial _q)]^2\chi _n(\xi )\varphi
_n(\vec q)\}.  \label{m19}
\end{equation}
In the equation (\ref{m18}) and (\ref{m19}) the functions $\chi _n(\xi )$
are flavour wave function of the quarks,
\[
\Gamma ^{(2)}(\vec q)=(\widehat{Q}_1^2+\widehat{Q}_2^2)(M^2-10E_q^2)\frac{
\vec E^2}{8E_q^4}+
\]
\[
+(\widehat{Q}_1^2+\widehat{Q}_2^2)(53E_q^2-5M^2)\frac{(\vec E\vec q)^2}{
16E_q^6}-3\widehat{Q}_1\widehat{Q}_2\frac{(\vec E\vec q)^2}{8E_q^4}.
\]
The wave functions $\varphi _n^{(1)}(\vec q,\xi )$ obey the quasipotential
equation in the first order in the external field. The interaction of the
dipole moment with the electric field yields the first term (\ref{m19}) $
(\alpha _{dip})$. All the other terms (\ref{m19}) are determined by the
relativistic interaction of the quarks with the external field $(\alpha
_{rel})$. The quasipotential equation (\ref{m16}) for the system of scalar
quarks in the first order in the field $\vec E$ is the form
\[
\lbrack {\vec \nabla _r}^2-\widetilde{V}(r)-m^2+\frac{M_n^2}2\frac{(\widehat{%
Q}_2-\widehat{Q}_1)}{2e}(e\vec E\vec r)]\varphi _n^{(1)}(\vec q,\xi )=0.
\]
We shall consider the model with Coulomb and oscillator quasipotential ($%
C=const$)
\[
V_{coul}=-\frac gr-C,
\]
\begin{equation}
V_{osc}=-gr^2-C.  \label{m20}
\end{equation}
The wave function for the quasipotential (\ref{m20}) have the form ($n=0$)
\[
\varphi _0^{coul}(r)=Aexp(-\beta _0^{coul}r),~~\varphi
_0^{osc}(r)=Aexp(-\frac 16\beta _0^{osc}r^2).
\]
The parameters of wave functions $\varphi _0(r)$ are determined by the
condition of the energy quantum
\[
\beta _n^{coul}=\frac{g^2}{4(n+1)^2},~~\beta _n^{osc}=-\sqrt{g}(n+\frac 32),
\]
where
\[
\beta _n^{coul}=m^2-C-\frac{M_n^2}4,~~\beta _n^{osc}=\frac{M_n^2}4-m^2+C.
\]
The parameters $g$ and $C$ are in agreement with the masses $\pi ^{^{\prime
}}(1300)$ and $K^{^{\prime }}(1400)$. For $\pi $-mesons $m=330MeV$, for $K$%
-mesons $m=420MeV$. Under these parameters the averaged squared radiuses of $%
\pi $- and $K$-mesons (\ref{m18}) have the same order in Coulomb and
oscillator models:
\[
\sqrt{\langle r_\pi ^2\rangle ^{coul}}=0.25f_m,~~\sqrt{\langle r_\pi
^2\rangle ^{osc}}=0.21f_m,
\]
\[
\sqrt{\langle r_K^2\rangle ^{coul}}=0.26f_m,~~\sqrt{\langle r_K^2\rangle
^{osc}}=0.23f_m.
\]
We obtain the same order for polarizability in Coulomb and oscillator models
(the numerical values are given in units of $10^{-4}f_m^3$).
\[
\alpha _{coul}^{\pi ^0}=1.83,~~\alpha _{coul}^{\pi ^{\pm }}=3.63,~~\alpha
_{coul}^{K^0}=0.24,~~\alpha _{coul}^{K^{\pm }}=0.93;
\]
\[
\alpha _{osc}^{\pi ^0}=1.68,~~\alpha _{osc}^{\pi ^{\pm }}=4.02,~~\alpha
_{osc}^{K^0}=0.27,~~\alpha _{osc}^{K^{\pm }}=1.05.
\]
Due to the factor $\frac 1M$ in (\ref{m19}), the relativistic correction to
the polarizability $(\alpha _{rel})$ make the essential contribution
compared with an ordinary polarizability $(\alpha _{dip})$.

\begin{center}
{\bf RQM formalism for $q\bar{q}$ bound states}
\end{center}

The formulation of relativistic quantum mechanics (RQM) differs from
nonrelativistic quantum mechanics by the replacement of invariance under
Galilean transformations with invariance Poincare transformations. The
dynamics of many-particle system in the RQM is specified by expressing the
ten generators of the Poincare group $\hat{M}_{\mu \nu }$ and $\hat{P}_\mu $
in terms of dynamical variables. In the constructing generators for
interacting systems it is customary to start with the generators of the
corresponding noninteracting system ({\bf we shall write this operators
without ''a hat'' }) and then add interactions in the a way that is
consistent with Poincare algebra. In the relativistic case it is necessary
to add an interaction $V$ to more than one generator in order to satisfy the
commutation relations of the Poincare algebra. Dirac \cite{Di1} observed,
that there is no unique way of separating the generators into dynamical
subset ( the generators including interaction $V$ ) and kinematical subset.
Kinematical subset must be associated with some subgroup Poincare
group,usually called stability group \cite{Leut} or kinematic subgroup \cite
{Coest1}.

There are three forms of the dynamics in the relativistic quantum mechanics,
called "instant","point","light-front" forms \cite{Di1}. The description in
the instant form implies that the operators of three-momentum and angular
momentum do not depend on interactions i.e. ${\hat{\vec P}}$ = ${\vec P}$
and ${\hat {\vec J}}$ = ${\vec J}$ ~~(${\hat {\vec J}}$=(${\hat M}^{23}$, ${
\hat M}^{31}$, ${\hat M}^{12}$)) and interactions may be presents in
operator $\hat P^0$ and generators of the Lorentz boosts ${\hat {\vec N}}=({
\hat M}^{01},{\hat M}^{02},{\hat M}^{03})$. The description in the point
form implies that the operators ${\hat M}^{\mu\nu}$ are the same as for
noninteracting particles, i.e.${\hat M}^{\mu\nu}= M^{\mu\nu}$, and thus
interaction terms can be present only in the four-momentum operators ${\hat P
}$. In the front form with the marked $z$ axis we introduce the + and -
components of the four-vectors as $p^+= (p^0+p^z)/\sqrt{2}$, $p^-=(p^0-p^z)/
\sqrt{2}$. We require that in the front form the operators ${\hat P}^+,{\hat
P}^j, {\hat M}^{12},{\hat M}^{+-}, {\hat M}^{+j}$ $(j=1,2)$ are the same as
the corresponding free operators, and interaction terms may be present in
the operators ${\hat M}^{-j}$ and ${\hat P}^-$. It is significant that
four-momentum of the two-particle bound system $P$ is not conserved i.e. $
P\ne p_1+p_2$, where $p_1$ and $p_2$ are four-momenta of the particles of
noninteracting system.

The three momenta $\vec{p}_1$, $\vec{p}_2$ of the particles (in our case,
the quarks ) with the masses $m_1$ and $m_2$ of relativistic system can be
transformed to the total $\vec{P}$ and relative momenta $\vec{k}$ to
facilitate the separation of the center mass motion:
\[
\vec{P}_{12}=\vec{p}_1+\vec{p}_2,
\]
\begin{equation}
\overrightarrow{k}=\overrightarrow{p_1}+\frac{\overrightarrow{P_{12}}}{M_0}%
\left( \frac{\left( \overrightarrow{P_{12}}\overrightarrow{p_1}\right) }{%
\omega _{M_0}\left( \overrightarrow{P_{12}}\right) +M_0}+\omega _{m_1}\left(
\overrightarrow{p_1}\right) \right) ,  \label{veck}
\end{equation}
where
\begin{equation}
M_0=\omega _{m_1}\left( \overrightarrow{k}\right) +\omega _{m_2}\left(
\overrightarrow{k}\right) ,\omega _{m_1}\left( \overrightarrow{p_1}\right) =%
\sqrt{\overrightarrow{p_1}^2+m_1^2}.  \label{mass}
\end{equation}

The eigenvalue problem for the mass of $q\bar q$ system can be expressed in
the two equivalent forms \cite{Pol1}:
\[
\hat M \mid \Psi> \equiv ( \hat M_{0} + \hat V ) \mid \Psi> = M \mid \Psi>,
\]
\begin{equation}
(k^{2}+\hat W) \mid\Psi> = \eta\mid\Psi>,
\end{equation}
where the mass of bound $q\bar q$ system $M$ and $\eta$ have relationship:
\[
M^{2}=2\eta+m^{2}_{1}+ m^{2}_{2}+2\sqrt{\eta(%
\eta+m^{2}_{1}+m^{2}_{2})+m^{2}_{1}m^{2}_{2}}
\]
and $k\equiv \left|\vec k\right|$.

The solution any of the above eigenvalue problems will lead to
eigenfunctions of the form
\[
_0\left\langle \vec{V}_{12}\mu ,\left[ J\hskip 2ptk\right] ,(ls)\right.
\left| \vec{V}\mu ,\left[ J\hskip 2ptM\right] \right\rangle =
\]
\begin{equation}
=\left( {\frac M{M_0}}\right) ^{3/2}\delta _{JJ^{\prime }}\delta _{\mu \mu
^{\prime }}\delta (\vec{V}-\vec{V}_{12})\Psi ^{J\mu }\left( k\hskip 2ptl%
\hskip 2pts\hskip 2pt;M\right)
\end{equation}
with the velocities of bound system $\vec{V}={\vec{P}}/{M}$ and
noninteracting system $\vec{V}_{12}={\vec{P}_{12}}/{M_0}$. The function $%
\Psi ^{J\mu }\left( k\hskip 2ptl\hskip 2pts\hskip 2pt;M\right) $ satisfies
in the point form a following equation \cite{Pol1}:
\[
\sum_{l^{\prime }s^{\prime }}\int\limits_0^\infty <k\hskip 2ptl\hskip %
2pts\parallel W^J\parallel k^{\prime }\hskip 2ptl^{\prime }\hskip %
2pts^{\prime }>\Psi ^J(k^{\prime }\hskip 2ptl^{\prime }\hskip 2pts^{\prime };%
\hskip 2ptM)k^{\prime 2}dk^{\prime }+
\]
\begin{equation}
+k^2\Psi ^J(k\hskip 2ptl\hskip 2pts;M)=\eta \Psi ^J(k\hskip 2ptl\hskip %
2pts;M)  \label{maineq}
\end{equation}
with reduced matrix element of operator $\hat{W}$.

The state vector $\left| \vec{V}_{12}\hskip 2pt \mu ,\hskip 2pt\left[
J\hskip 2pt k\right] ,\hskip 2pt (l\hskip
2pts)\right\rangle _0$ is the eigenstate of operators $\vec{V}_{12}$, $\hat{{%
J}^2}$ (angular momentum), $\hat{J}_3$ and also $\hat{{L}^2}$, $\hat{{S}^2}$
, where $\vec{L}$ and $\vec{S}$ are relative orbital momentum and total spin
momentum accordingly. This vector of the noninteracting $q\bar{q}$ system
transforms irreducibly under Poincare transformations. The vector $\left|
\vec{V}\hskip 2pt\mu ,\left[ J\hskip 2pt M\right]
\right\rangle $ is eigenstate of the interacting system, that also
transforms irreducibly.

In this approach the meson state is defined by as state of on-shell quark
and antiquark with the wave function $\Psi ^{J\mu }\left( k\hskip 2ptl\hskip %
2pts\hskip
2pt;M\right) $
\begin{eqnarray}
\left| \overrightarrow{P}\mu \hskip 2pt\left[ JM\right] \right\rangle
&=&\sum_{ls}\sum_{\lambda _1\lambda _2}\int d^3k\sqrt{\frac{\omega
_{m_1}\left( \overrightarrow{p_1}\right) \omega _{m_2}\left( \overrightarrow{
p_2}\right) M_0}{\omega _{m_1}\left( \overrightarrow{k}\right) \omega
_{m_2}\left( \overrightarrow{k}\right) \omega _{M_0}\left( \overrightarrow{
P_{12}}\right) }}  \nonumber \\
&&\ \Psi ^{J\mu }\left( kls;M\right) \sum_{m\lambda }\sum_{\nu _1\nu
_2}\left\langle s_1\nu _1,s_2\nu _2\right| \left. s\lambda \right\rangle
\left\langle lm,s\lambda \right| \left. J\mu \right\rangle  \nonumber
\\
&&\ Y_{lm}\left( \theta ,\phi \right) D_{\lambda _1\nu _1}^{1/2}\left(
\overrightarrow{n}\left( p_1,P_{12}\right) \right) D_{\lambda _2\nu
_2}^{1/2}\left( \overrightarrow{n}\left( p_2,P_{12}\right) \right)  \nonumber
\\
&&\left| p_1\lambda _1\right\rangle \left| p_2\lambda _2\right\rangle
\label{state}
\end{eqnarray}
where $\left\langle s_1\nu _1,s_2\nu _2\right| \left. s\lambda \right\rangle
$, $\left\langle lm,s\lambda \right| \left. J\mu \right\rangle $ are
Clebsh-Gordan coefficients of $SU(2)$-group, $Y_{lm}(\theta ,\phi )$ -
spherical harmonic with spherical angle of $\vec k$. Also, in Eq.(\ref{state}
)
\[
D^{1/2}\left( \overrightarrow{n}\right) =\frac{1-i\left( \overrightarrow{n}
\overrightarrow{\sigma }\right) }{\sqrt{1+\overrightarrow{n}^2}}
\]
is $D$-function of Wigner rotation, which determined by vector-parameter
\[
\overrightarrow{n}\left( p_1,p_2\right) =\frac{\overrightarrow{u_1}\times
\overrightarrow{u_2}}{1-\left( \overrightarrow{u_1}\overrightarrow{u_2}
\right) }
\]
with $\overrightarrow{u}=\overrightarrow{p}/\left( \omega _m\left(
\overrightarrow{p}\right) +m\right) $.

\begin{center}
{\bf Basic Requirements of Current Operator}
\end{center}

Current operators $\hat{J}^\mu (x)$ of bound systems is required for a
calculation of the decay constants, the charge form factors and the other
properties of relativistic particles. Since $\hat{J}^\mu (x)$ is a
four-vector operators, it has same transformation properties as the
four-momentum $\hat{P}_\mu $ under Poincare transformation. It implies that
the commutation relations between $\hat{J}^\mu (x)$ and Poincare generators $
\hat{M} ^{\rho \sigma }$, $\hat{P}_\mu $ is identical to the commutation
relations between the generators of the Poincare group and four momentum:
\begin{equation}  \label{tok3}
\left[ \hat{M}^{\rho \sigma },\hat{J}^\mu (x)\right] =i\left( g^{\mu \sigma }
\hat{J}^\rho \left( x\right) -g^{\mu \rho } \hat{J}^\sigma \left( x\right)
\right) -i\left( x^\rho \frac{\partial \hat{J}^\mu \left( x\right) }{
\partial x_\sigma }-x^\sigma \frac{ \partial \hat{J}^\mu \left( x\right) }{
\partial x_\rho }\right)
\end{equation}
\begin{equation}  \label{tok4}
\left[ \hat{P}_\mu ,\hat{J}_\gamma (x)\right] =-i\frac{ \partial \hat{J}_\mu
\left( x\right) }{\partial x^\gamma }
\end{equation}

The translational invariance of the current operators implies that
\begin{equation}  \label{tok5}
\hat{J}_\mu \left( x\right) =\exp \left( i\hat{P} x\right) \hat{J}_\mu
\left( 0\right) \exp \left( -i\hat{P}x\right)
\end{equation}
This relation makes it possible to reduce problem of seeking $\hat{J}_\mu
\left( x\right) $ to the problem of seeking $\hat{J}_\mu (0)$ (see, for
example \cite{Lev2}, \cite{Keist1} ). The requirements of Lorentz invariance
(\ref{tok3}) reduces to
\begin{equation}  \label{tok6}
\left[ \hat{M}^{\rho \sigma },\hat{J}^\mu (0)\right] =i\left( g^{\mu \sigma }
\hat{J}^\rho \left( 0\right) -g^{\mu \rho } \hat{J}^\sigma \left( 0\right)
\right)
\end{equation}
If the theory is invariant under the space reflection and time reversal, and
${\hat U}_P$,${\hat U}_T$ are the corresponding representation operators,
then the current operator should be satisfy the following conditions:
\begin{eqnarray}
&&{\hat U}_P({\hat J}^0(x^0,{\vec x}),{\hat {\vec J}}(x^0,{\vec x})) {\hat U}
_P^{-1}=({\hat J}^0(x^0,-{\vec x}),-{\hat {\vec J}}(x^0,-{\vec x})),
\nonumber \\
&&{\hat U}_T({\hat J}^0(x^0,{\vec x}),{\hat {\vec J}}(x^0,{\vec x})) {\hat U}
_T^{-1}=({\hat J}^0(-x^0,{\vec x}),-{\hat {\vec J}}(-x^0,{\vec x}))
\label{ptime}
\end{eqnarray}
In addition to these equations continuity equation $\partial \hat{J}^\mu
\left( x\right) /\partial x^\mu $ $=0$ can be used. As follows from (\ref
{tok4}) this requirements can be written in the form
\begin{equation}  \label{tok8}
\left[ \hat{P}_\mu ,\hat{J}_\nu (0)\right] g^{\mu \nu }=0.
\end{equation}
Finally, the operator $\hat{J}_\mu \left( x\right) $ should be also satisfy
the cluster separability \cite{Pol1,Keist1,Lev4,Lev5} for many-particle
system.

For calculations many authors assumed that the mathematical expressions of
current operators for bound system and for the noninteracting system are
equal. This condition ( so called, impulse approximation )
\begin{equation}  \label{tok7}
\hat{J}_\mu \left( 0\right) =J_\mu (0)
\end{equation}
reasonable only in point form of the Poincare relativistic mechanics (\cite
{Lev3}). This result is evident from Eq.(\ref{tok6}).

\begin{center}
{\bf Decay constants and charge radius of pseudoscalar mesons in RQM
formalism }
\end{center}

The decay constant $f_p$ for $\pi ^{\pm }$ is defined by
\begin{equation}  \label{dec2}
\left\langle 0\left| \hat{J}^\mu \left( 0\right) \right| \overrightarrow{P}
,M\right\rangle =i\left( 1/2\pi \right) ^{3/2}\frac 1{\sqrt{2\omega _M\left(
\overrightarrow{P}\right) }}P^\mu f_p,  \label{deconst}
\end{equation}
where $\hat{J}^\mu (0)$ is the operator axial-vector part of the charged
weak current after a Cabibbo-Kobayashi-Maskawa mixing matrix element $
V_{qq^{\prime }}$ has been removed. The state vector is normalized by $
\left\langle \vec{P},M\right. \left| \vec{P}^{\prime },M\right\rangle $ =$
\delta (\vec{P}-\vec{P}^{\prime })$.

The current operator of the meson can be defined by the axial-vector current
of the free quarks fields:
\[
\hat{J}^\mu \left( 0\right) ={\hat{d}}\left( 0\right) \gamma ^\mu \gamma _5{
\hat{u}}\left( 0\right) .
\]
Using Eq.(\ref{state}) and Eq.(\ref{dec2}) we found, that
\begin{equation}
f_p=\frac{\sqrt{2}m\hskip 2pt N_c}{\pi \sqrt{M}}\int_0^\infty \frac{dk
\hskip 2pt k^2}{\omega _m\left( \overrightarrow{k}\right) }\Psi
\left( k,M\right) ,  \label{dec1}
\end{equation}
where $N_c$-number of colors and the wave function for pseudoscalar meson
have normalization condition

\[
\int_0^\infty dk\hskip 2pt k^2\hskip 2pt N_c \left|
\Psi \left( k,M\right) \right| ^2=1
\]
For pion we assumed, that the masses of $u$ and $d$-quarks are equal i.e. $
m_u=m_d\equiv m$. In the nonrelativistic quark model, when $k^2\ll m^2$,
meson decay constants are given
\begin{equation}
f_p\sim \frac 1{\sqrt{M}}\Psi \left( r=0\right)  \label{nonrel}
\end{equation}
with $\Psi \left( r=0\right) \sim \int d^3k\hskip 2pt\Psi (k)$.

Amplitude for decay $\pi \to \gamma +\gamma $ can be parameterized as
\begin{equation}
M_{\pi ^0\rightarrow \gamma \gamma }(k_1,k_2)=(-1)\frac{4\pi \alpha }{\left(
2\pi \right) ^{3/2}}\hskip 2ptg_{\pi \gamma \gamma }\hskip 2pt\epsilon _{\mu
\nu \rho \sigma }\hskip 2ptk_1^\mu \hskip 2ptk_2^\nu \hskip 2pt\xi ^\rho
\left( k_1\right) \hskip 2pt\xi ^\sigma \left( k_2\right) ,  \label{dec3}
\end{equation}
where $k_1$,\hskip 2 pt $k_2$ are momenta of photons having polarization
vectors $\xi ^\rho \left( k_1\right) $ and $\xi ^\rho \left( k_2\right) $.
The calculations of the matrix element $M_{\pi ^0\rightarrow \gamma \gamma
}(k_1,k_2)$ for real photons in the lowest order implies that
\begin{equation}
g_{\pi \gamma \gamma }=\frac{m\hskip 2ptN_c}{\pi \hskip 2ptM^{3/2}}
\int_0^\infty \frac{dy\hskip 2pty}{\sqrt{y^2+1}}\ln \left| \frac{\sqrt{y^2+1}
+y}{\sqrt{y^2+1}-y}\right| \Psi \left( y\hskip 2ptm,M\right)  \label{dec4}
\end{equation}

The equation of motion of the bound $q\bar q$ states (\ref{maineq}) in the
RQM is relativistic equation with appropriate effective potential $W$ (or $V$
). However, it is hard problem to obtain wave function $\Psi(k,M)$ as
solution of this equation. Therefore, we use simple model wave function
depending on length scale parameter $\beta$:
\begin{equation}
\Psi (k \hskip 2pt,M)\equiv \Psi (k \hskip 2pt,\beta)=2/(\sqrt{N_{c}}
\beta^{3/2} \pi^{1/4}) exp(-\frac{k^{2}}{2\beta^{2}}).  \label{wavef}
\end{equation}

And second parameter of this model is the constituent quark mass $m$. They
are fixed by fitting the relevant experimental data. We use the following
values for decay constant $f_p=130.7\pm 0.1\pm 0.36~~MeV$~\cite{Pdg1}~ and
for coupling constant $g_{\pi \gamma \gamma }=0.0839\pm 0.0013~~MeV^{-1}$,
which determined from radiative widths $\Gamma (\pi \to \gamma +\gamma )$= $
7.25\pm 0.13~~eV$~\cite{Pgg}~. From these values we obtain the allowed
region for the mass $m$ of quark and the parameter $\beta $ in wave function
~(\ref{wavef}):
\begin{equation}
m=155\pm 5~~~MeV,~~~~\beta =150\pm 5~~~MeV.
\end{equation}
Our optimal value of light quark mass $m=155~~MeV$ is not in agreement with
result of $m=250~~MeV$ (see,~\cite{Jaus1}~, ~\cite{Ito}~), but is roughly in
agreement with result of $m=170~MeV$ ~\cite{Munz1}~.

The most general form of the hadronic matrix element of the vector current
must be represented in terms of two form factors

\begin{eqnarray*}
\left\langle \overrightarrow{P^{\prime }},M^{\prime }\right| \overline{q}
^{\prime }(0)\gamma ^\mu q(0)\left| \overrightarrow{P,}M\right\rangle
&=&\frac 1{2*(2\pi )^3\sqrt{\omega _M\left( \overrightarrow{P}\right) \omega
_{M^{\prime }}\left( \overrightarrow{P^{\prime }}\right) }} \\
&&(F_{+}\left( t\right) \left( P\mu ^{\prime }+P_\mu \right) +F_{-}\left(
t\right) \left( P_\mu ^{\prime }-P_\mu \right) ),
\end{eqnarray*}

where $t=-\left( P^{\prime }-P\right) ^2,$ $q(x)-$ operator of the quark
field. In the case of a pion we shall consider only $F_{+}\left( t\right) $.
On rearrangement, as a result we obtain following expression for the form
factor:

\begin{eqnarray*}
F_{+}\left( t\right) &=&\frac 12\int\limits_0^\infty
dk k^2 \Psi \left( k\right)
\int\limits_0^1 dx [\Psi \left( k_1\right) +\Psi \left(
k_2\right) - \\ &&-\frac{k \left| \overrightarrow{V}\right|
}{\omega _m\left( \overrightarrow{k}\right) } x \left(
\Psi \left( k_1\right) -\Psi \left( k_2\right) \right) ],
\end{eqnarray*}

where $k_i^2=k^2+\Delta _i,$ $\Delta _i=4\gamma ^4[\left| \overrightarrow{V}
\right| ^2\left( \omega _m^2\left( \overrightarrow{k}\right) +k^2x^2\right)
+(-1)^i\left| \overrightarrow{V}\right| k$ $(1+\left| \overrightarrow{V}
\right| ^2),$ $\gamma =1/\left( 1-\left| \overrightarrow{V}\right| ^2\right)
,$ $\left| \overrightarrow{V}\right| ^2=t/(t+4$ $M^2).$

Using the definition mean square radius : $\left\langle r^2\right\rangle =-6$
$dF_{+}\left( t\right) /dt$ at $t=0$ we obtain the final relativistic pion
charge radius

for the wave function (\ref{wavef}) in simple form:

\begin{equation}
\left\langle r^2\right\rangle =\frac 3{M^2}\left( \frac{m^2}{2\beta
^2}+\frac 54\right) .  \label{rad}
\end{equation}

Unfortunately for authors, numerical calculation of the mean square radius
is not in agreement with experimental result $\left\langle r^2\right\rangle
=0.44$ $fm^2$ for pion. But we thinks, that more appropriate form of the
wave function $\Psi \left( k\right) $ furnishes the desired result.

\begin{center}
{\bf ACKNOWLEDGMENTS}
\end{center}

This work was supported by grant {\bf N. F96-326 (17.02.97}) from the
Foundation for Fundamental Research of Republic Belarus.


\begin{thebibliography}{99}
\bibitem{BeteS}  N.Nakanichi, Progr. Theor. Phys. Suppl., N43 (1969);
E.E.Salpeter,H.A.Bethe, Phys. Rev. {\bf 84 }, p.232 (1951).

\bibitem{LogTav}  A.A.Logunov, A.N.Tavkhelidze, Nuovo Cimento,{\bf \ 29},
p.380 (1963).

\bibitem{Kad1}  V.G.Kadyshevsky, Nucl.Phys. B, {\bf 6 }p.125 (1968).

\bibitem{Tod}  I.T.Todorov, Phys.Rev. D,{\bf \ 10, }p.2351 (1971).

\bibitem{Gross}  F.Gross , Phys.Rev.,{\bf \ 186}, p.1448 (1969); Phys. Rev.
D, {\bf 10}, p.223 (1974).

\bibitem{Keist1}  B.D.Keister, W.N.Polyzou in Advances in Nuclear Physics,
edited J.W.Negele and E.Vogt (Plenum, New York, 1991).

\bibitem{Lev4}  F.M.Lev Annals of Physics {\bf 237}, p.355 (1995).

\bibitem{Wein}  S.Weinberg, Phys.Rev. {\bf 150}, p.1313 (1966).

\bibitem{FrS}  L.L.Frankfurt, M.I.Strikman Phys.Rep., {\bf 76}, p.215 (1981).

\bibitem{KonTer}  L.A.Kondratyuk, M.V.Terent'ev Yad.Fiz., {\bf 31}, p.1087
(1980).

\bibitem{Kad2}  V.G.Kadyshevsky, Zh. Eksp. Teor. Fiz., {\bf 46}, p.654,
p.872 (1964); V.G.Kadyshevsky, R.M.Mir-Kasimov, N.B.Skachkov Yad. Fiz., {\bf %
9},p.219, p.462 (1969).

\bibitem{Kad3}  V.G.Kadyshevsky, R.M.Mir-Kasimov, N.B.Skachkov Part. Nucl.,
{\bf 2}, p.635 (1972).

\bibitem{m2}  R.N.Faustov, Nuovo Cimento.,{\bf 69A}, p.37-46, (1970);
M.K.Volkov. Part. and Nuclei, {\bf 24}, p.81 (1993); T.Ino. Progr. Theor.
Phys., {\bf 89}, p.895 (1993).

\bibitem{m3}  N.V.Maksimenko, S.G.Shulga, Yad. Fiz., {\bf 56}, p.201 (1993).

\bibitem{m5}  N.V.Maksimenko, L.G.Moroz, In Proc. 2 Intern. School of Young
Scientists on High Energy Physics and Relativistic Nucl. Phys., Gomel, 1979.
Dubna JINR,D2-11707, p.533-543 1979.

\bibitem{m6}  A.I.L'vov, Preprint FIAN. N344, Moscow. 1987.

\bibitem{m7}  N.V.Maksimenko, L.G.Moroz, O.M.Deryuzhkova. Proceeding of the
Academy of Sciences of Belarus, Phys-math. series, 1995, N4, p.118-122

\bibitem{m8}  A.A.Arkhipov,V.I.Savrin, Fiz. Elem. Chast. and Atom. Yad.
(Sov. J. Part. Nucl.), {\bf 16}. p.1091 (1985).

\bibitem{m9}  C.Itzykson, J.B.Zuber. Quantum field theory. Moskow:
''Peace'', v.1,2. 1984 (In russian).

\bibitem{m10}  D.Lurie, A.Macfarlane, Y.Takahashi., Phys. Rev., {\bf 140B},
p.1091 (1965).

\bibitem{m11}  S.R.Amendolia et. al. , Phys. Lett. , {\bf B178}, p.435
(1986).

\bibitem{m12}  Yu.M.Antipov et. al., Z. Phys. ,{\bf C26}. p.495 (1985).

\bibitem{Di1}  P.A.M. Dirac Rev.Mod.Phys., {\bf 21}, p.392 (1949)

\bibitem{Leut}  H. Leutwyler and J. Stern, Annals of Phys. ${\bf 112}$,94
(1978).

\bibitem{Coest1}  F. Coester and W.N. Polyzou, Phys.Rev. D, ${\bf 26}$,1348,
(1982).

\bibitem{Pol1}  W.N. Polyzou, Annals of Physics {\bf 193}, p.367 (1989).

\bibitem{Lev2}  F.M.Lev ''Forms of Relativistic dynamics, current operators
and deep inelastic lepton-nucleon scattering'' hep-ph/9505373.

\bibitem{Lev5}  F.M.Lev , Rivista del Nuovo Cimento, {\bf 16, }N2 , p.1
(1993).

\bibitem{Lev3}  F.M. Lev '' Effect of binding in deep in\-elastic
scat\-tering revisited '', hep-ph/9501348.

\bibitem{Pdg1}  Review of Particle Properties, Phys. Rev.D, {\bf 50}, p.1271
(1994).

\bibitem{Pgg}  H.Atherton et al. Phys. Lett. B .{\bf 158}, p.1 (1985).

\bibitem{Jaus1}  W.Jaus Phys.Rev. D, {\bf 44}, N.9, p.2851 (1991)

\bibitem{Ito}  H.Ito, W.W.Buck and F.Gross Phys. Rev.C, {\bf 45}, p.1918
(1992)

\bibitem{Munz1}  C.R. M\"unz, J. Resag, B.C. Metsch and H.R. Petry Phys.Rev.
C {\bf 52} N.4, p.2110 (1995).
\end{thebibliography}
\end{document}